# Nitrogen-doped carbon quantum dots enhance bacteria growth and biosensing signals

David Rutherford[1], Marketa Šlapal Bařinková[1], Jaroslav Kuliček[1], Jelena Kozic[2], Jovana Prekodravac Filipović[2], Bohuslav Rezek[1]

[1] *Faculty of Electrical Engineering, Czech Technical University in Prague, Prague, Czech Republic*
[2] *Vinča Institute of Nuclear Sciences-National Institute of the Republic of Serbia, University of Belgrade, Serbia*

## Abstract

Carbon quantum dots (CQDs) are known for their antibacterial properties and ability to inhibit bacteria growth. In the current study, we observed a dopant and concentration-dependency on the antibacterial effect of CQDs. High concentrations of CQDs completely inhibited bacteria growth yet low concentrations enhanced growth. Unlike undoped CQDs, nitrogen-doped CQDs (N-CQDs) enhanced bacteria growth in a concentration-dependant manner and nitrogen/iron co-doped CQDs (Fe/N-CQDs) resulted in growth profiles similar to untreated bacteria. N-CQDs also exhibited the strongest photoluminescence (PL) signal which was quenched by bacteria, and the reduction in the maximum PL intensity was linear over the concentration range tested until saturation. N-CQDs were found to be located around the bacteria cell periphery suggesting intimate interactions. Illumination prior to interaction with bacteria had little influence on these growth effects. Absorbance measurements of colloidal CQD, N-CQD and Fe/N-CQD confirmed long-term stability (7 days). Such materials have potential for incorporation into rapid sensing and diagnostic systems for bacteria detection in liquids for biomedical applications.

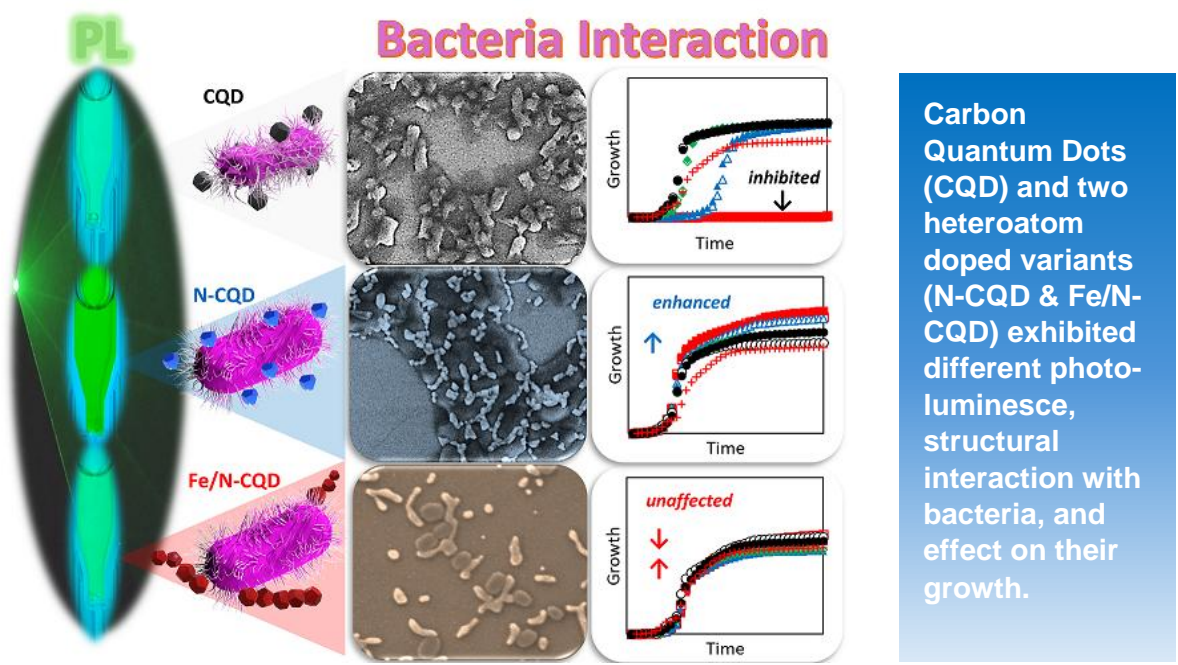

# 1. Introduction

Carbon quantum dots (CQDs) are nanomaterials derived from carbon, typically less than 10 nm in diameter, which possess unique optical and electronic properties due to quantum mechanical effects[1–3]. CQDs have been successfully synthesized by various techniques, and the synthesis environment heavily influences their properties[4,5]. A common optoelectronic property of CQDs irrespective of the synthesis technique is strong photoluminescence (PL) due to the quantum confinement effect and the relationship between a particles dimension and exciton Bohr radius. The PL signal is tuneable depending on the particle size: smaller quantum dots emit blue light whilst larger quantum dots emit red light[6,7]. CQDs absorb a broad spectrum of light with the majority in the UV region, but emit a very narrow, specific wavelength, making them ideal candidates for precise applications such as bioimaging and biosensing[8]. The surface of CQDs can also be rich in functional groups that can be tuned to further improve their performance in specific applications like biosensing, bioimaging or conjugation[9].

It has been revealed that heteroatom doping of CQDs can improve the optoelectronic properties by adjusting the bandgap and introducing new energy levels[8]. Dopants can create mid-gap states that leads to new emission wavelengths that can be used for targeted sensing applications. Doped CQDs with specific luminescent properties serve as excellent contrast agents for bioimaging[10] or for detecting antibiotics[11], heavy metals[11,12] and even reactive oxygen species[13] by the changes in optical or electronic properties. Nitrogen-doped carbon quantum dots (N-CQDs) are a special type of CQD where nitrogen atoms are incorporated into the carbon lattice. This doping enhances the optical, electronic, and chemical characteristics of CQDs, making them more suitable for a variety of advanced biosensing applications. N-doping enhances the PL quantum yield through the introduction of new energy states (e.g., n -> π* transitions), improving the quantum yield and brightness of N-CQDs[14]. N-CQDs often exhibit tuneable fluorescence emission across the visible spectrum and the incorporation of N-containing functional groups (e.g. amino, pyrrolic or graphitic nitrogen) improves the water solubility of N-CQDs, making them suitable for bioimaging in aqueous environments[15].

CQDs have unique physicochemical properties such as PL, high surface area, and functional capability, which enables interaction with bacteria in diverse ways. Due to their small size, disruption to the bacterial membranes and internalization leads to interference with normal cell functions that underpin the antibacterial mechanism similar to other nano-sized materials[16]. However, depending on the CQD surface properties, it is also possible to stimulate bacteria growth. It has been recently reported that selenium-doped CQDs (Se-CQDs) increased microbial diversity and abundance which lead to an enhancement of plant growth[17]. It was postulated that the greater hydrophilicity and content of Se-CQDs were responsible for the growth-promoting effects. Yet interestingly, CQDs synthesized by the same method but from different precursors induced an antibacterial response[18]. Elsewhere, graphene oxide quantum dots (GOQDs) improved the proliferation of *B. cereus* compared to graphene oxide flakes[19]. These studies highlight the complexity of the interaction between QDs and bacteria which depends greatly on the specific QD properties. The typical trend of high concentrations of the smallest nanoparticle inducing the optimal antibacterial effect does not always apply to CQDs. The reason for this observed dichotomy is not well understood and in the current study we attempt to explore the less studied phenomena of bacterial growth promotion in response to CQD exposure. Characterisation of CQD surface properties in conjunction with analysis of growth kinetic parameters such as length of lag phase and maximum growth rate of bacteria grown in the presence of different CQD concentrations offer an opportunity to study the mechanisms that are responsible for the enhancement of bacteria growth.

# 2. Methods

## 2.1. QD synthesis, characterisation, and pre-illumination

### 2.1.1. Microwave-based synthesis of CQD, N-CQD and Fe/N-CQD

Quantum dots consisting of carbon (CQD), nitrogen-doped carbon (N-CQD) and iron-nitrogen co-doped carbon (Fe/N-CQD) were all synthesized using microwave (MW)-assisted heating technique. The scheme is shown in Figure 1. A clear glucose solution (0.1 g/mL, D-Glucose monohydrate, Fluca) was placed inside a professional MW reactor (Anton Paar Monowave 300) for 15 min, which caused a change in colour to deep brown indicating the successful synthesis of CQDs. This type of MW reactor provides more specific and precise control of the reaction conditions compared to domestic MW ovens. Solutions were dialyzed for 5 days (3.5 kDa) and filtered using a vacuum pump (KNF Laboport) through different pore size filter membranes from 250 to 10 nm (Polycarbonate membrane disk, GVS Filter Technology, USA).

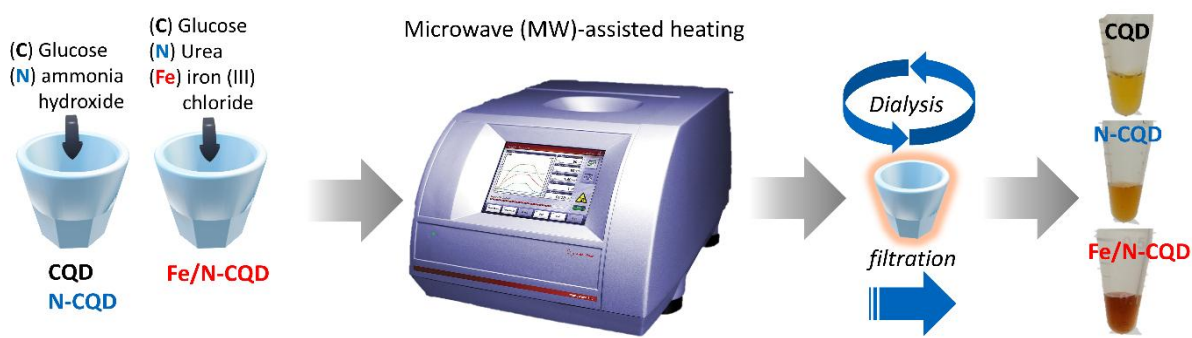

*Figure 1 Schematic summary of the microwave-assisted synthesis technique used to prepare CQD, N-CQD and Fe/N-CQD. Coloured letters in brackets beside the chemicals denote the element the precursor was used for.*

Synthesis of N-CQDs was performed using the same MW-assisted method as described above but for 1 min instead of 15 min[20]. Starting materials were a 0.1 g/mL glucose aqueous solution and ammonium hydroxide (25%) in a 5:1 volume ratio. Solutions were dialyzed and filtered as stated above for CQDs.

The Fe/N-CQDs were prepared by stirring the glucose (0.1 g/ml), urea (0.005 g/ml) and iron (III) chloride trihydrate ($FeCl_3$ x $3H_2O$) (0.055 g/ml) water solution for 2 h, followed by MW irradiation under the same conditions as N-CQDs[20]. Solutions were again dialyzed and filtered as above and all synthesized QDs were stored at room temperature until needed. A stock colloidal solution of each QD sample was prepared by weighing 15 mg of dry powder and subsequently adding 5 mL of HPLC $H_2O$ (concentration = 3 mg/mL).

### 2.1.2. Photoluminescence (PL)

PL spectra for CQD samples were obtained using a confocal Photoluminescence microscope (WITec alpha300 RAS) equipped with a white light quasi-continuum tunable laser (NTK Photonics, excitation wavelength 532 nm, laser power 100 µW). PL spectra were collected by a 100x objective (NA 0.9) and UHTS 300 VIS spectrometer. The integration time was 30 s and the number of accumulations was 3. A spectral longpass filter (550 nm) was put before the detector to filter the signal from the laser. Spectra were collected and processed by WITec Control 6.1 and WITec Project 6.1 software. A 1:10 dilution from the stock suspensions was performed using HPLC $H_2O$ before dropcasting 5 µL onto a pre-cleaned Si substrate for analysis.

Quenching of the PL signal in response to the addition of bacteria was measured at first using a single bacteria concentration for all QD samples. This was then repeated for a range of bacteria concentrations for N-CQDs only. More information on the bacteria strain used and preparation is provided in section **Error! Reference source not found.**. Equal volumes of colloidal CQDs (1:10 dilution from stock) and bacteria in water were incubated for 1 hour (37 °C) before drop-casting 5 µL onto pre-cleaned Si substrate for analysis. PL spectra acquisition settings as described above.

### 2.1.3. Absorbance

Optical absorbance profiles of CQD colloids were measured using UV-vis spectroscopy (200-900 nm, 2 nm resolution, Epoch 2, Biotek). Individual wells of a UV-transparent 96-well plate (pureGrade S, Brand) containing 200 µL of 150 µg/mL of each CQD colloid was measured in triplicate. Colloidal stability was assessed by monitoring the change in absorbance for 7 consecutive days (absorbance peak = 230 nm for carbon). All colloids were analysed as prepared and were not subjected to further ultrasonication treatment.

### 2.1.4. Zeta (ζ-) potential

The surface charge of the CQDs was determined by ζ-potential measurement using a zetasizer (Zetasizer, Malvern Panalytical Instruments, UK). The pH of each colloidal suspension was measured immediately after zeta potential measurement by a pH sensor of the Zetasizer titrator unit.

## 2.2. CQD-bacteria interaction

### 2.2.1. Bacteria preparation

Bacteria preparation involved the creation of vials containing an overnight culture of *Escherichia coli* (*E. coli*, ATCC 25922, CCM) mixed with glycerol (2:1) and stored at -20 °C before use. For each experiment, a single vial (3 mL) was thawed and serially diluted using 900 µL of 0.9 % NaCl (Penta) and 100 µL of bacteria suspension before inoculating 500 µL of each dilution onto Mueller Hinton agar (MHA, Sigma) contained in 60 mm Petri dishes. After overnight incubation at 37 °C, a single colony was removed, added to 5 mL Mueller Hinton broth (MHB, Sigma) contained in a 15 mL falcon tube placed and on an orbital shaker (150 rpm, 20 h) located inside an incubator (37 °C). The resulting clonal populations were adjusted to McFarland's Density 1.0 which is equivalent to $3 \times 10^8$ colony forming units per millilitre (cfu/mL) before being further diluted 1000x using MHB so that the initial bacteria concentration of the experiments ([cell]0) is approximately $3 \times 10^5$ cfu/mL.

### 2.2.2. Pre-illumination of colloids

Pre-illumination of CQD colloids in wells of a UV-transparent microwell plate was performed using solar simulated light (HAL-C100, Asahi Spectra). Four different concentrations of each CQD were uniformly exposed to the light source operating at 100% for 30 minutes (distance from light source to QD = 40 cm). The same concentrations of each CQD not exposed to solar simulated light were then added to different wells for comparison.

### 2.2.3. Bacteria Growth

Bacteria in MHB (50 μL) were added to the appropriate wells of a 96-well microplate that contained either pre-illuminated QD colloids, unilluminated colloids, or only water (i.e., positive control). The optical density (OD) of the incubated (37 °C) bacteria-CQD samples were recorded every 30 minutes over a 20 h period and the initial OD value was used to blank-correct the data. Growth parameters were then calculated from these data: the length of the lag phase ($t_{lag}$) was determined as the time when the OD value became greater than 0.2 absorbance units, the maximum growth rate (μ) was the value recorded for the greatest difference in OD per unit time ($\Delta_{OD}/\Delta_t$) and the time when this occurred ($\mu_t$). The final cell concentration ([cell]$_{final}$) was determined by standard broth dilution technique and addition to MHA for overnight cultivation (37 °C) before imaging the bacteria growth in the petri dishes the following day using an automatic colony counter (SphereFlash).

### 2.2.4. Scanning Electron Microscopy (SEM)

The interaction between CQD colloids and bacteria was visualized using SEM (Evo 10, Zeiss). Equal volumes of CQD colloids and bacteria in water were mixed by orbital rotation at 37 °C for 1 hour before dropcasting 5 μL onto a pre-cleaned silicon substrate for analysis without fixation or sputter coating. The acceleration voltage (5 kV) and working distance (8.5 mm) were kept constant for each sample and imaged using the secondary electron (SE) emission detection mode.

### 2.2.5. Statistical analysis

One-way analysis of variance (ANOVA) was used to test the significance among differences in the growth profile parameters and [cell]$_{final}$ after exposure to different concentrations of each CQDs (α = 0.05). Post hoc two-tailed t-tests were used to assess the significance of the differences between CQD groups.

# 3. Results and Discussion

## 3.1. Colloidal stability

The surface potential of colloidal CQDs, N-CQDs and Fe/N-CQDs in water was estimated from ζ-potential measurements and revealed a relatively weak negative charge for all CQDs. CQDs recorded a value of -19.3 ± 1.0 mV (pH 5.79), followed by Fe/N-CQDs with a value of -13.5 ± 1.2 mV (pH = 6.31) and the least negative were N-CQDs with the value of -8.4 ± 1.1 mV (pH 7.04). Generally, high ζ-potential values (i.e., > ± 30 mV) indicate good colloidal stability due to enhanced electrostatic repulsions between particles that leads to the greater electrostatic potential over the solvent. Our measured values would infer less than optimal colloidal stability

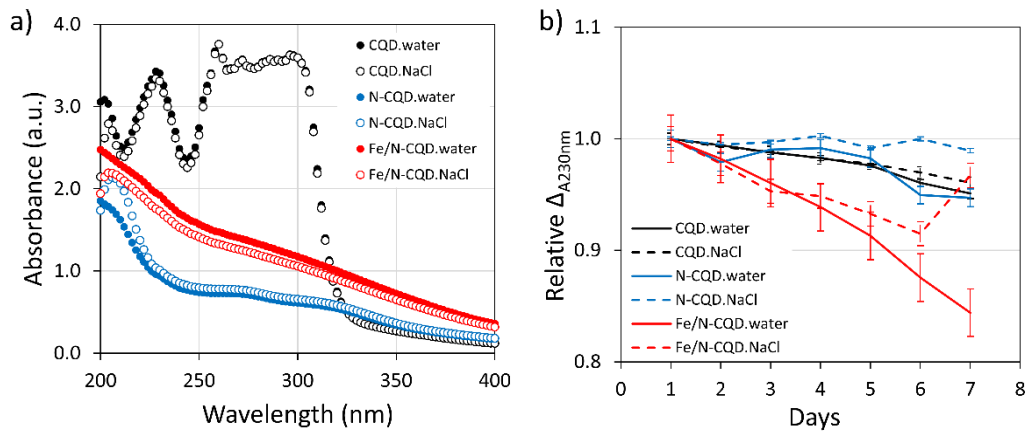

*Figure 2 a) UV absorbance spectra (200-400 nm) for 150 µg/mL of CQD, N-CQD and Fe/N-CQD in water and 0.9% NaCl, and b) shows the change in absorbance value at 230 nm relative to the initial absorbance value ($\Delta_{A230nm}$) over 7 days. Error bars show the standard deviation of the mean (n=3).*

In addition to ζ-potential measurements, we assessed the long-term stability of CQDs in both liquid types using UV-vis spectroscopy. All CQD colloids were analysed as prepared without additional ultrasonication treatment that is typically required when using nanoparticle-based colloids. The absorbance profiles can be seen in **Error! Reference source not found.**a and the maximum absorbance peak of carbon at approximately 230 nm was visible from the CQD spectra which is attributed to π-π* transitions of aromatic sp2 domains of the carbon core[14,21,22]. Whilst the solvent is known to affect the absorbance profile of colloidal CQDs[22], we recorded similar intensities in both water and NaCl. The absorbance peak was absent from N-CQD and Fe/N-CQD spectra in both liquid types which suggests surface modification or passivation due to the presence of dopants which has also been observed elsewhere[23,24].

Monitoring the change in the position and intensity of the absorbance peak maxima of colloidal CQDs over time was used to confirm long-term optical absorbance capability[25]. The absorbance value at 230 nm ($A_{230nm}$) was selected to assess the stability of the synthesised CQD colloids by monitoring the change in intensity over time and the results are presented in **Error! Reference source not found.**b. N-CQDs in NaCl recorded the smallest reduction (1 %) in $A_{230nm}$ over 7 days whereas a greater reduction (5 %) was observed in water. The same trend was also observed for CQDs and Fe/N-CQDs and the greatest reduction in $A_{230nm}$ over time was recorded in water compared to NaCl, except for the final measurement of Fe/N-CQDs in NaCl that showed an unexpected increase from day 6 to day 7. The type of liquid and complexity would influence the colloidal stability of CQDs and hence the interaction with co-suspended bacteria. All CQD colloids showed good stability over time in both liquid types and there was little difference in the stability profiles in water compared to the more ionically complex 0.9% NaCl. Prior study of the luminescence signal intensity from CQDs to assess the stability of the colloid in NaCl was comparable to the current study[26]. It was not possible to study CQD stability in MHB due to the carbon-rich biomolecules contained within the liquid which saturated the absorbance signal in the same (UV) range.

$A_{230nm}$ measurements over 7 days suggest relatively stable colloids, in particular N-CQDs. However, the relatively weak ζ-potential measurements imply otherwise and N-CQDs recorded the least negative value. The discrepancy can be explained by how the different techniques assess colloidal stability. ζ-potential measures the effective surface charge on the particle relative to the surrounding bulk liquid and the magnitude of the charge is associated with agglomeration prevention. $A_{230nm}$ measurement is typically used to monitor the change in position or intensity of the peak absorbance maxima and any significant variation from the initial value would indicate colloidal instability.

## 3.2. Photoluminescence (PL)

Optical images of CQDs taken prior to photoluminescence (PL) analysis show different agglomeration characteristics and interaction with bacteria (**Error! Reference source not found.**a). A halo was observed around bacteria that were exposed to CQDs and considerable particle localization around bacteria was seen for N-CQDs. Fe/N-CQDs formed long chains of agglomerates and were less localized around bacteria. PL properties of the synthesized CQDs and the influence bacteria addition had on the PL signal intensity can be seen in **Error! Reference source not found.**b. The strongest PL signal was produced by N-CQDs without bacteria and displayed two distinct peaks at 640 nm and 675 nm which were also observed in the CQD and Fe/N-CQD spectra but at much lower magnitudes (see inset **Error! Reference source not found.**b). This enhancement in PL intensity is in good agreement with other reports which showed an increase in PL intensity from CQDs doped with N[27]. The presence of bacteria did not change the position of either peak but clearly influenced the PL signal intensity. Since the strongest PL signal was produced by N-CQDs, this material was chosen for further studies involving different concentrations of bacteria. It was shown that the intensity of the PL signal of both peaks reduced with increasing bacteria concentration, until no further reduction was possible (**Error! Reference source not found.**c).

Excitation using green light (i.e., 532 nm laser) resulted in emission of light in the infrared (IR) and near-infra red (NIR) regions of the spectrum. Previous analysis of colloidal N-CQDs and Fe/N-CQDs using a range of shorter excitation wavelengths (350-490 nm) resulted in the emission of a single peak maxima located at shorter wavelength positions relative to the current analysis[14]. The wavelength position of the PL maximum red-shifted as the excitation wavelength increased, which is in good agreement with other studies that reported a similar effect with CQDs[28,29]. This can explain the reason for the different PL spectra from the same CQD colloids when we used different excitation wavelengths.

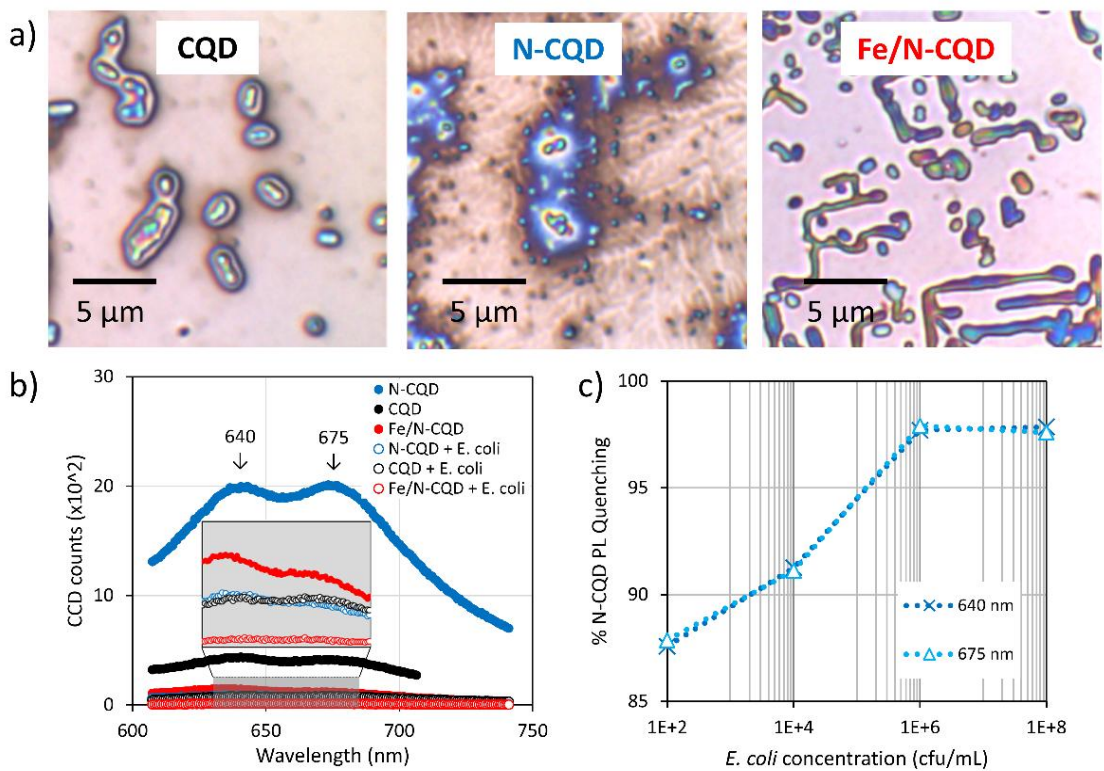

*Figure 3 a) optical images (100x) of CQDs with E. coli. b) Photoluminescence spectra (with the substrate signal subtracted) of CQDs with E. coli (10^8 cfu/mL). c) N-CQD PL signal quenched (%) upon interaction with a range of concentrations of E. coli (10^2-10^8 cfu/mL).*

Monitoring the change in PL signal from CQDs can be used in a range of biosensing applications. An array of CQDs modified with antibiotics was previously utilized as a fluorescence-based detector of foodborne pathogens from both water and meat[30]. It was demonstrated that monitoring the change in fluorescence intensity after incubation with the bacteria was a rapid and sensitive method for detection. It was further revealed that different bacteria species caused variation in the degree of fluorescence quenching, which also adds specificity to the technique. Non-destructive detection and tracking of live Gram-positive bacteria was achieved using CQDs through observations of peptidoglycan-mediated quenching of the fluorescence signal[31].

Moreover, Gram-negative *E. coli* exhibited negligible response when compared to the untreated bacteria, suggesting a method for differentiating Gram-positive bacteria from Gram-negative bacteria without performing the Gram Stain technique. A reduction in PL signal intensity as a result of polypeptide or amino acid-mediated quenching from bacteria culture liquid and reactive oxygen species has also been reported[32].

## 3.3. Bacteria growth

The effect colloidal CQDs had on bacteria growth was assessed by monitoring the change in optical density (OD) over time and measuring the concentration of bacteria at the end of the experiment. The OD at the start of the measurement for each CQD type and concentration was used for blank correction for the remaining data of that sample so that the OD measurements were not influenced by the co-suspended CQDs. The complete (blank-corrected) growth curves are provided in Figure S1. The OD data was used to extract characteristic growth parameters such as the length of the lag phase ($t_{lag}$), maximum growth rate ($\mu$) and the time the maximum growth rate was recorded ($\mu_t$). These results are summarised in Figure 4 as a function of CQD concentration. In addition, the final cell concentration ([cell]$_{final}$) was also measured from the pre-illuminated CQD samples.

It's important to note that presence of colloidal nanoparticles also contributes to absolute OD value of the bacteria suspension. This was the reasoning behind using the initial OD value for blank correction. Nevertheless, the change in OD at 600 nm from each CQD type in MHB over the equivalent growth period of bacteria growth was measured to assess any CQD effect on OD values throughout the incubation time. The results indicated minimal effects on the overall OD values and the subsequent influence on the OD values during bacteria growth was negligible (Figure S2). The highest concentration of N-CQDs recorded an increase after 20 h of only 0.025 absorbance units, which is equivalent to less than 2 % of the maximum final OD value recorded (Figure S3). We are therefore able to categorically state that any differences observed in the OD values are overwhelmingly due to differences in how bacteria have grown and not the presence of co-suspended CQDs.

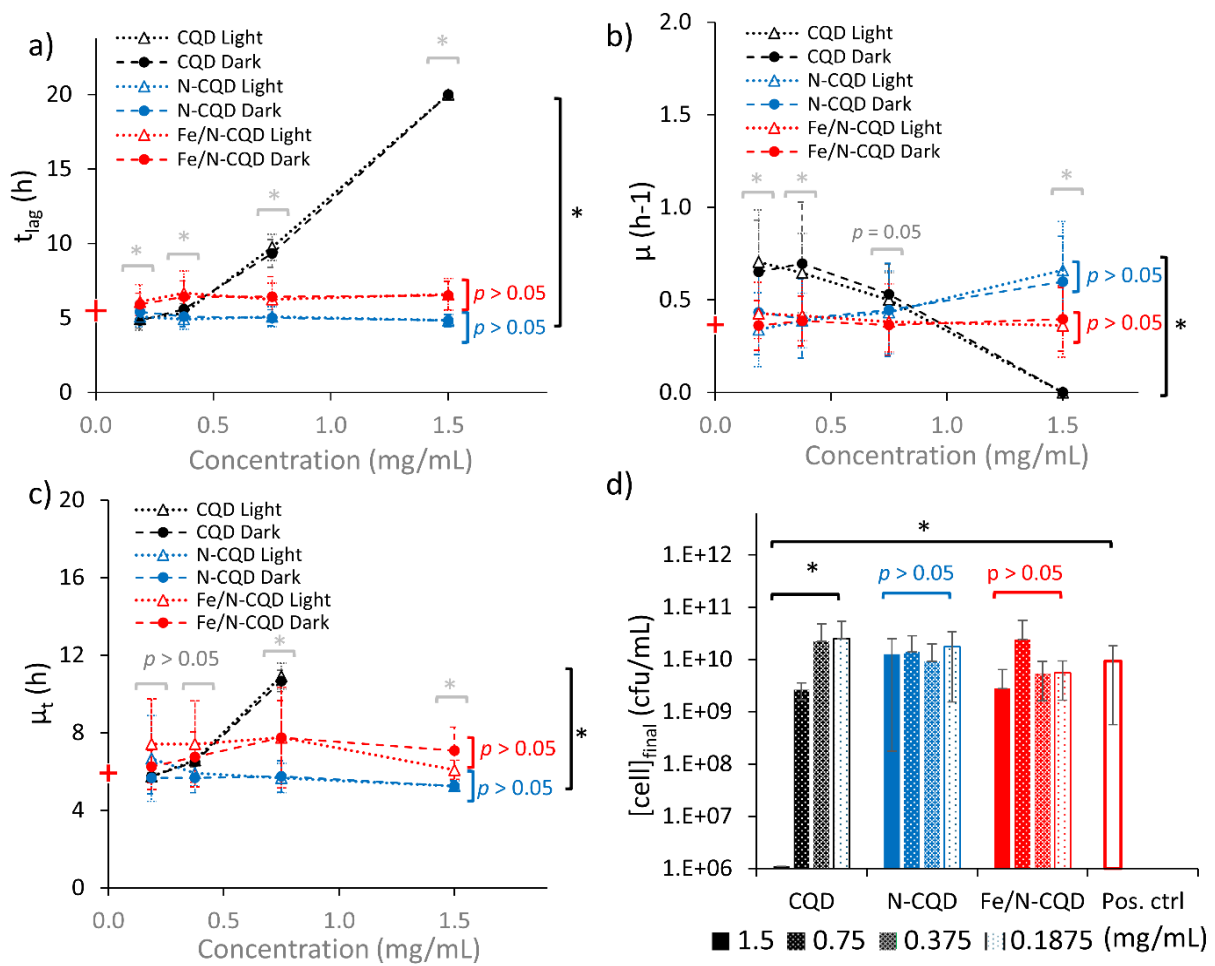

*Figure 4 Growth parameters extrapolated from optical density (OD) measurements of bacteria growth with different concentrations of CQD (Figure S2). a) Length of lag phase ($t_{lag}$) shown as a function of CQD concentration for all test samples. b) Maximum growth rate ($\mu$) shown as a function of CQD concentration for all test samples. c) The time during growth when the maximum growth rate was recorded ($\mu_t$) as a function of quantum dot concentration. d). Data points are the mean and error bars show the standard deviation of the mean (n = 6). The final bacteria cell concentrations ([cell]$_{final}$) recovered after 20 h exposure to different concentrations of QD (numbers in legend refer to QD concentration in mg/mL). Error bars are standard deviations of the mean (n = 3) and \* = p < 0.05.*

### 3.3.1. Lag Phase duration

Bacteria growth during exposure to CQDs was negatively affected in a concentration-dependent manner and follows the typical trend of a nanomaterial which possesses antibacterial properties (Figure 4a). $t_{lag}$ increased with CQD concentration, and there was no statistically significant difference between CQDs that had been pre-illuminated with solar simulated light compared to CQDs without pre-illumination across the concentration range tested. The greatest CQD concentration completely inhibited bacteria growth (i.e., $t_{lag}$ = 20 h) and as the concentration reduced the inhibitory effect lessened as seen by the reduction in $t_{lag}$ closer to values recorded by untreated bacteria (positive control, red cross, Figure 4a). $t_{lag}$ for both N-CQD and Fe/N-CQD were not influenced by the concentration and the differences were not significant across the concentration range investigated regardless of pre-illumination. However, the differences in $t_{lag}$ recorded over this concentration range of CQDs was statistically significant. For a given concentration, the differences in $t_{lag}$ between CQD types were statistically significant except for pre-illuminated CQD ($t_{lag}$= 4.9 h) and both N-CQD samples (light = 5.1 h; dark = 5.42 h).

The increase in $t_{lag}$ in a concentration-dependant manner observed here is in good agreement with other studies that exposed *E. coli* to CQDs during growth. CQDs synthesized via the hydrothermal method with opposite surface charges were co-cultured with *E. coli* and it was revealed that the positively-charged CQDs produced a greater increase in $t_{lag}$ due to electrostatic attraction with the negatively-charged bacteria cell[33]. In the current work, all synthesized CQDs were negatively charged therefore the differences we observe in bacteria growth are not due to electrostatic interaction with the negatively charged bacteria cell. EDA-capped CQDs also extended the $t_{lag}$ of *E. coli* post-treatment and samples that were pre-illuminated with visible light produced a longer $t_{lag}$ than without pre-illumination[34]. The distance from bacteria to light source was not described there. In our experiments the distance was 40 cm to uniformly illuminate all samples at once which could be the reason why we did not observe any photo-induced effects on $t_{lag}$.

### 3.3.2. Maximum growth rate

CQD exposure resulted in a reduction in $\mu$ with concentration, which is typical of an antibacterial effect, and the differences in $\mu$ over the concentration range tested were statistically significant (**Error! Reference source not found.**b). Whilst the highest CQD concentration inhibited growth, the lowest concentration enhanced growth and resulted in the fastest $\mu$ all samples. Enhancement of bacteria growth after exposure to sub-lethal concentrations of nanoparticles has also been reported elsewhere[35]. A vastly different growth response was observed when bacteria were exposed to N-CQDs, where $\mu$ increased with increasing N-CQD concentration. Furthermore, the greatest $\mu$ was recorded from the highest concentration of N-CQD that were pre-illuminated, which suggests a possible photo-catalytic enhancement of growth. However, for lower N-CQD concentrations, $\mu$ for pre-illuminated samples was less than the unilluminated counterparts. Interestingly, there was an opposite trend recorded in $\mu$ as a function of Fe/N-CQD concentration depending on pre-illumination. Bacteria exposed to Fe/N-CQDs without pre-illumination recorded a steady increase in $\mu$ with concentration, whereas there was a surprising initial increase in $\mu$ for low concentrations of pre-illuminated Fe/N-CQDs before a gradual reduction. Nevertheless, the differences observed in $\mu$ for *E. coli* exposed to a range of N-CQD and Fe/N-CQD concentrations were not statistically significant.

Other studies of *E. coli* exposed to CQDs produced growth profiles identical to unexposed bacteria, however, the concentration range assessed was lower (0-200 µg/mL) than the current study and did not show any variation in µ[36]. Elsewhere, µ was negatively impacted in a concentration-dependant manner when *E. coli* were exposed to sub-inhibitory concentrations of AuNPs and AgNPs[37–39]. This suggests that any decrease in µ is associated with the nanomaterial that interacts with the bacteria having antibacterial properties and triggering an antibacterial response.

### 3.3.3. Time of maximum growth rate

$\mu_t$ increased with increasing CQD concentration (Figure 4c), which correlates well with $t_{lag}$ and µ trends that indicate antibacterial effects. Bacteria exposed to Fe/N-CQD without pre-illumination recorded similar $\mu_t$ to untreated bacteria, however there was a consistent reduction in $\mu_t$ for bacteria exposed to Fe/N-CQD with pre-illumination. A further reduction in $\mu_t$ was observed from bacteria exposed to N-CQD that remained consistent irrespective of concentration or pre-illumination. The differences observed in $\mu_t$ after exposure to a range of CQD concentrations was significant, however exposure to the same concentrations of N-CQD or Fe/N-CQD resulted in differences that were not significant.

### 3.3.4. Final cell concentration

The initial cell concentration was the same for all experiments, standardised to approximately $3 \times 10^5$ cfu/mL. Figure 4d shows the concentration of bacteria that were measured at the end of the experiment ([cell]$_{final}$) and one can see that the highest concentration of CQD inhibited normal bacteria growth and was below the detection limit of $10^6$ cfu/mL. The difference between [cell]$_{final}$ after exposure to 1.5 mg/mL CQD and the remaining CQD concentrations was statistically significant. The difference between [cell]$_{final}$ from the positive control and [cell]$_{final}$ exposed to 1.5 mg/mL CQDs was statistically significant. Lower CQD concentrations recorded a slightly greater [cell]$_{final}$ relative to the control, except for the lowest concentration which resulted in a slight (0.5 log) reduction however these differences were not statistically significant. Therefore, the faster µ recorded from the lowest concentration of CQDs did not translate into a significantly greater [cell]$_{final}$. Bacteria exposed to N-CQD grew to similar [cell]$_{final}$ concentrations as the untreated bacteria irrespective of concentration and the difference was not statistically significant. Finally, bacteria that were exposed to Fe/N-CQD during growth recorded slightly lower [cell]$_{final}$ concentrations, except for 0.75 mg/mL that resulted in a slight increase but the overall differences in [cell]$_{final}$ were not statistically significant.

In summary, CQDs resulted in a concentration-dependant increase in $t_{lag}$ and $\mu_t$, with a reduction in µ that are typical traits of an antibacterial material, although even the highest concentration assessed still did not completely inhibit bacteria growth (1.5 mg/mL). The opposite trends were observed using N-CQD, where $t_{lag}$ and $\mu_t$ were reduced relative to untreated bacteria and µ increased which suggests a growth promoting effect. Bacteria exposed to Fe/N-CQD recorded slightly greater $t_{lag}$ compared to untreated bacteria, yet the effect on $\mu_t$ and µ varied depending on pre-illumination. These results clearly show that doped QD can have quite different effects on bacteria growth compared to undoped QD, and in the case of N-doped CQD the complete opposite effect was observed.

Whilst there are many reports in the literature on the antibacterial effect of CQDs, highlighting their suitability for novel antibacterial treatments to combat AMR[16,33,34,40,41], our results show that the antibacterial properties of CQDs, in particular N-CQDs, are not universal. There were few prior studies that also showed no significant difference in [cell]$_{final}$ of *E. coli* after exposure to CQDs[36] suggesting that there were no adverse effects on bacteria growth in response to CQD exposure. N-CQDs synthesized using the solvothermal method also did not reduce the concentration of *E. coli* for similar exposure times and concentrations used in the current study[42]. N-CQDs were shown even to enhance digestion under anaerobic conditions to promote bacteria growth[9,43]. Such material-enhanced hybrid biocatalysis systems (MHBS) can boost microbial metabolism by influencing mass and electron transfer in the extracellular, interfacial and intracellular environments[44]. These mechanisms might also be at play in our closed culture system once the oxygen has been depleted from within the wells of the microwell plate used for bacteria growth.

### 3.3.5. Interaction with bacteria

To better understand the above effects, attempts were made to image the bacteria with QD directly from the antibacterial test. Unfortunately, crystallization of broth components and metabolites prevented clear visualisation of the interaction in that environment. Therefore, a parallel study was performed in water to mimic the interaction between CQD and bacteria after 1 hour incubation (37°C). The results can be seen in Figure 5 and clearly reveal different interaction behaviour. Bacteria and CQD (Figure 5a) are irregularly shaped and show a high degree of damage, whereas the bacteria with N-CQD (Figure 5b) and Fe/N-CQD (Figure 5c) appear less damaged with the majority of membranes intact. N-CQD formed small agglomerates that encapsulated the bacteria and showed closer interaction than Fe/N-CQD which formed larger agglomerates not localised beside the bacteria. This microscopic view after bacteria-CQD interactions correlates very well with the bacteria growth curves and the observed effects on the characteristic growth parameters.

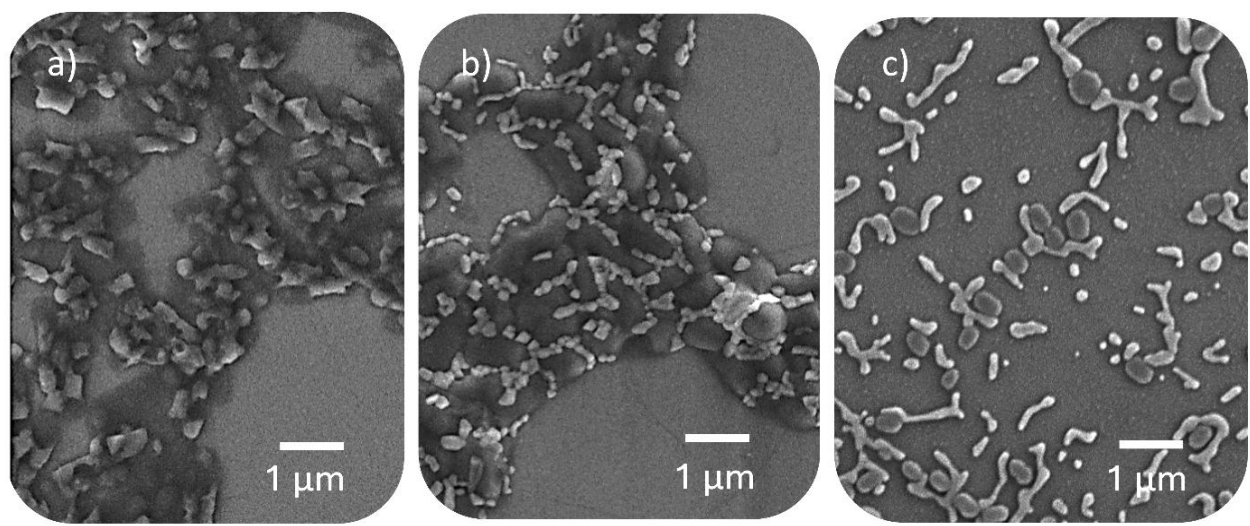

*Figure 5 SEM images of the interaction between bacteria and a) CQD, b) N-CQD and c) Fe/N-CQD. Full images of each sample available in the S.I.*

Previously, variation in the outer membrane morphology of *E. coli* was observed after exposure to fluorescent carbon dots (F-CD), where the smooth walls of the typical rod-shaped bacteria became severely roughened and damaged[45]. The mode of action was direct attachment of the F-CD onto the bacteria surface that resulted in membrane disintegration and lead to bacteria inactivation. CQDs and N-CQDs synthesized from a glucose precursor using the hydrothermal method displayed varying degrees of bacteria damage[46]. *E. coli* growth was more severely impacted by N-CQD exposure than by CQD and displayed a greater degree of morphological damage consisting of large-scare perturbations and cytoplasmic leakage around the cell periphery. In contrast, *E. coli* exposed to N-CQDs synthesized from anthracene displayed no adverse effects on viability with only slight morphological changes and no signs of cell rupture[42].

These data, in combination with the effects on bacteria growth, clearly highlights a discrepancy in the literature regarding CQD interactions with bacteria. Therefore, it is of paramount importance to thoroughly characterise all synthesized CQDs by multiple spectroscopic and microscopic analyses if the true mechanism of action in complex systems such as their interaction bacteria is to be fully understood.

# 4. Conclusion

The objective of this work was to explore the influence various CQDs prepared by the same synthesis technique on bacteria growth in real-time using standard microbiological growth parameters such as the length of the lag phase, maximum growth rate and the time when maximum growth rate occurred. We observed that N-CQDs promoted bacteria growth, and the highest concentration resulted in the maximum growth rate without varying the lag phase duration or time at which the maximum growth rate occurred. N-CQDs also produced the largest PL intensity that remained detectable in the presence of bacteria. High magnification images of bacteria-CQD interaction revealed N-CQD localisation around viable bacteria cells. Quite different behaviour on bacterial growth was observed from CQDs which showed a typical antibacterial response of increased lag phase duration, reduced maximum growth rate and increased time when the maximum growth rate occurred. CQDs produced similar PL spectra to N-CQDs but at much lower intensities and the bacteria showed extensive membrane damage after the interaction. Unlike N-CQDs and CQDs, the effect Fe/N-CQDs had on bacteria growth varied due to pre-illumination (by solar simulator). Lower concentrations of pre-illuminated Fe/N-CQDs resulted in a faster maximum growth rate compared to without pre-illumination, yet the effect was reversed for the maximum concentration tested. The PL signal intensity from Fe/N-CQD was much lower compared to that of N-CQDs and formed islands of agglomerates which were less localised around the bacteria cells and unsuitable for potential biosensing applications. These findings clearly show that CQDs can affect bacteria growth in vastly diverse ways depending on the presence of dopants, which also significantly alter the optoelectronic properties. We observed an inverse dependence between the concentration of bacteria and the intensity of PL signal generated from the least bactericidal N-CQD which is promising for future incorporation within novel bacteria detection systems.

# Acknowledgements

DR and JaK acknowledge the support of TAČR (TM03000033), MŠB and BR acknowledge the support of European Union and the Ministry of Education, Youth and Sports of the Czech Republic project No. CZ.02.01.01/00/22_008/0004596 (SenDiSo). MŠB would also like to acknowledge the student grant from CTU (SGS23/166/OHK4/3T/13). JPF and JK acknowledge the support of Science Fund of the Republic of Serbia grant No. #7741955 (PHO-TOGUN4MICROBES) and the Ministry of Science, Technological Development and Innovation of the Republic of Serbia grant No. 451-03-136/2025-03/200017.

# CRediT authorship contribution summary

**DR**: conceptualization, methodology, investigation, data curation & formal analysis (CQD absorbance, ZP, SEM, bacteria growth), writing (original & final version)
**MŠB**: data curation & formal analysis (bacteria growth), writing (review & editing)
**JaK**: investigation, data curation & formal analysis (PL)
**JK**: CQD synthesis and modification, investigation
**JPF**: conceptualization, methodology, data curation (CQD synthesis)
**BR**: funding acquisition, formal analysis, writing (review & editing)

# Data availability

Data for this article (zeta potential and dynamic light scattering, UV-vis spectroscopy, Photoluminescence and bacteria growth) are available at Zenodo [10.5281/zenodo.15130105].

## References

1      R. Jelinek, *Carbon Quantum Dots*, Springer International Publishing, Cham, 2017.
2      V. S. Sivasankarapillai, A. V. Kirthi, M. Akksadha, S. Indu, U. D. Dharshini, J. Pushpamalar and L. Karthik, *Nanoscale Adv.*, 2020, **2**, 1760–1773.
3      D. Ozyurt, M. A. Kobaisi, R. K. Hocking and B. Fox, *Carbon Trends*, 2023, **12**, 100276.
4      V. Magesh, A. K. Sundramoorthy and D. Ganapathy, *Front. Mater.*, DOI:10.3389/fmats.2022.906838.
5      L. Zhang, X. Yang, Z. Yin and L. Sun, *Luminescence*, 2022, **37**, 1612–1638.
6      H. Li, X. He, Z. Kang, H. Huang, Y. Liu, J. Liu, S. Lian, C. H. A. Tsang, X. Yang and S.-T. Lee, *Angewandte Chemie International Edition*, 2010, **49**, 4430–4434.
7      P. Van Duong, L. Anh Thi, L. Duc Toan, P. Hong Minh, N. Thanh Binh, D. Hoang Tung, N. Dinh Lam and N. Minh Hoa, *ChemistrySelect*, 2024, **9**, e202401754.
8      V. L. John, Y. Nair and T. P. Vinod, *Particle & Particle Systems Characterization*, 2021, **38**, 2100170.
9      Q. Liu, S. Chen, H. Zhang, L. Feng, H. Zhou, J. Pan, Y. Li and C. Xu, *Chemical Engineering Journal*, 2023, **475**, 146359.
10     M. J. Molaei, *Sci Rep*, 2022, **12**, 17681.
11     F. Wang, Y. Zhang, H. Li, W. Gong, J. Han, S. Jiang, D. Li and Z. Yao, *Food Chemistry*, 2025, **463**, 141122.
12     Q. Liu, B. Ren, K. Xie, Y. Yan, R. Liu, S. Lv, Q. He, B. Yang and L. Li, *Nanoscale Adv.*, 2021, **3**, 805–811.
13     X. Geng, T. R. Congdon, P. Anees, A. A. Greschner, F. Vetrone and M. A. Gauthier, *Nanoscale Adv.*, 2020, **2**, 4024–4033.
14     J. R. Prekodravac, B. R. Vasiljević, J. Žakula, M. Č. Popović, V. Pavlović, G. Ciasca, S. Romanò and B. M. Todorović Marković, *Optical Materials*, 2024, **147**, 114629.
15     A. Prakash, S. Yadav, P. Tiwari, P. S. Saxena, A. Srivastava and R. Tilak, *J Nanopart Res*, 2023, **25**, 237.
16     M. Yu, P. Li, R. Huang, C. Xu, S. Zhang, Y. Wang, X. Gong and X. Xing, *J. Mater. Chem. B*, 2023, **11**, 734–754.
17     K. Yin, Q. Bao, J. Li, M. Wang, F. Wang, B. Sun, Y. Gong and F. Lian, *Chemosphere*, 2024, **364**, 143175.
18     M. Saikia, A. Hazarika, K. Roy, P. Khare, A. Dihingia, R. Konwar and B. K. Saikia, *Journal of Environmental Chemical Engineering*, 2023, **11**, 111344.
19     L. Mu, Q. Zhou, Y. Zhao, X. Liu and X. Hu, *Journal of Hazardous Materials*, 2019, **366**, 694–702.
20     J. Prekodravac, B. Vasiljević, Z. Marković, D. Jovanović, D. Kleut, Z. Špitalský, M. Mičušík, M. Danko, D. Bajuk–Bogdanović and B. Todorović–Marković, *Ceramics International*, 2019, **45**, 17006–17013.
21     X.-W. Fang, H. Chang, T. Wu, C.-H. Yeh, F.-L. Hsiao, T.-S. Ko, C.-L. Hsieh, M.-Y. Wu and Y.-W. Lin, *ACS Omega*, 2024, **9**, 23573–23583.

22    M. S. Zaini, J. Y. C. Liew, S. Paiman, T. S. Tee and M. A. Kamarudin, *J Fluoresc*, 2025, **35**, 245–256.

23    Y. Liu, W. Li, P. Wu, C. Ma, X. Wu, M. Xu, S. Luo, Z. Xu and S. Liu, *Sensors and Actuators B: Chemical*, 2019, **281**, 34–43.

24    A. Aygun, I. Cobas, R. N. Elhouda Tiri and F. Sen, *RSC Advances*, 2024, **14**, 10814–10825.

25    X. Huang, .

26    W. U. Khan, D. Wang and Y. Wang, *Inorg. Chem.*, 2018, **57**, 15229–15239.

27    C. D. Stachurski, S. M. Click, K. D. Wolfe, S. Dervishogullari, S. J. Rosenthal, G. K. Jennings and D. E. Cliffel, *Nanoscale Adv.*, 2020, **2**, 3375–3383.

28    D. Carolan, C. Rocks, D. B. Padmanaban, P. Maguire, V. Svrcek and D. Mariotti, *Sustainable Energy Fuels*, 2017, **1**, 1611–1619.

29    X.-D. Mai, Y. T. H. Phan and V.-Q. Nguyen, *Advances in Materials Science and Engineering*, 2020, **2020**, 9643168.

30    M. Xiao, L. Mei, J. Qi, L. Zhu and F. Wang, *Microchemical Journal*, 2024, **201**, 110701.

31    C. Yan, C. Wang, T. Hou, P. Guan, Y. Qiao, L. Guo, Y. Teng, X. Hu and H. Wu, *ACS Appl. Mater. Interfaces*, 2021, **13**, 1277–1287.

32    Z. Lu, C. M. Li, H. Bao, Y. Qiao and Q. Bao, *Journal of Nanoscience and Nanotechnology*, 2009, **9**, 3252–3255.

33    X. Hao, L. Huang, C. Zhao, S. Chen, W. Lin, Y. Lin, L. Zhang, A. Sun, C. Miao, X. Lin, M. Chen and S. Weng, *Materials Science and Engineering: C*, 2021, **123**, 111971.

34    M. J. Meziani, X. Dong, L. Zhu, L. P. Jones, G. E. LeCroy, F. Yang, S. Wang, P. Wang, Y. Zhao, L. Yang, R. A. Tripp and Y.-P. Sun, *ACS Appl. Mater. Interfaces*, 2016, **8**, 10761–10766.

35    D. Rutherford, J. Jíra, K. Kolářová, I. Matolínová, J. Mičová, Z. Remeš and B. Rezek, *Int J Nanomedicine*, 2021, **16**, 3541–3554.

36    S. Qiang, L. Zhang, Z. Li, J. Liang, P. Li, J. Song, K. Guo, Z. Wang and Q. Fan, *Antioxidants*, 2022, **11**, 2475.

37    X. Zhang, X. Wang, H. Cheng, Y. Zheng, J. Zhao and K. Qu, *Journal of Hazardous Materials*, 2021, **413**, 125320.

38    M. L. Navarro-Pérez, M. C. Fernández-Calderón and V. Vadillo-Rodríguez, *Applied and Environmental Microbiology*, DOI:10.1128/aem.01849-21.

39    X. Zhang, Q. Yang, L. Ma, D. Zhang, W. Lin, N. Schlensky, H. Cheng, Y. Zheng, X. Luo, C. Ding, Y. Zhang, X. Hou, F. Lu, H. Yan, R. Wang, C.-Z. Li and K. Qu, *Biosensors and Bioelectronics*, 2023, **239**, 115626.

40    N. A. Travlou, D. A. Giannakoudakis, M. Algarra, A. M. Labella, E. Rodríguez-Castellón and T. J. Bandosz, *Carbon*, 2018, **135**, 104–111.

41    P. Li, F. Han, W. Cao, G. Zhang, J. Li, J. Zhou, X. Gong, G. Turnbull, W. Shu, L. Xia, B. Fang, X. Xing and B. Li, *Applied Materials Today*, 2020, **19**, 100601.

42    Y.-Y. Zheng, K.-T. Huang, S.-J. Lee, J.-S. Ni and Y.-H. Hsueh, *Process Biochemistry*, 2024, **146**, 225–233.

43    Q. Liu, H. Zhang, H. Wang, S. Zhao, Y. Li, L. Feng, J. Pan, H. Zhou and C. Xu, *Chemical Engineering Journal*, 2024, **500**, 156792.

44    H. Peng, Y. Su, X. Fan, S. Wang, Q. Zhang and Y. Chen, *Water Research*, 2025, **268**, 122759.

45    J. Liang, W. Li, J. Chen, X. Huang, Y. Liu, X. Zhang, W. Shu, B. Lei and H. Zhang, *ACS Appl. Bio Mater.*, 2021, **4**, 6937–6945.

46    P. Ezati, J.-W. Rhim, R. Molaei, R. Priyadarshi, S. Roy, S. Min, Y. H. Kim, S.-G. Lee and S. Han, *Sustainable Materials and Technologies*, 2022, **32**, e00397.

# Supporting Information

## Growth curves (ΔOD/t)

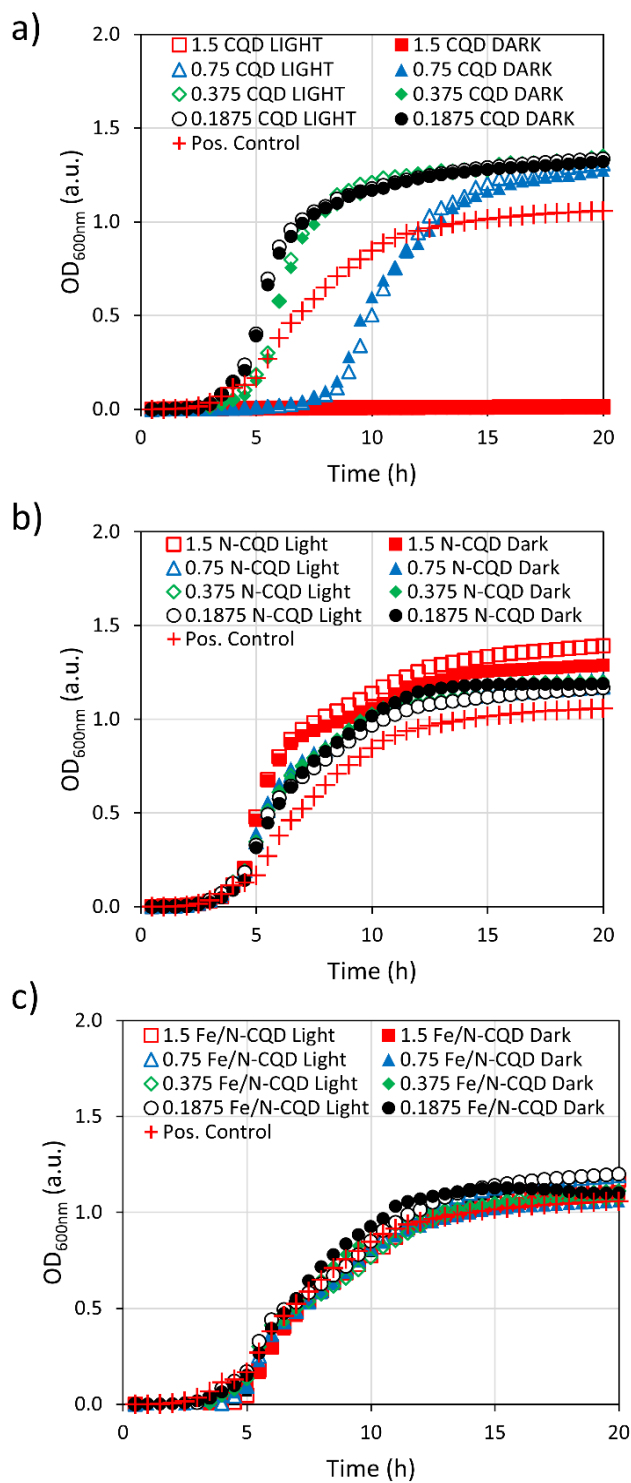

**Figure S1** *The change in optical density at 600 nm ($OD_{600\,nm}$) over time of bacteria exposed to 1.5 mg/mL, 0.75 mg/mL, 0.375 mg/mL and 0.1875 mg/mL CQDs. 'Light' denotes pre-illuminated material, whereas 'dark' is for samples without pre-illumination. Positive control shows bacteria growth in the absence of quantum dots. All data points are averaged (mean) values (n=6 for samples and n=12 for positive control).*

# Optical density (OD) values of CQDs only (no bact)

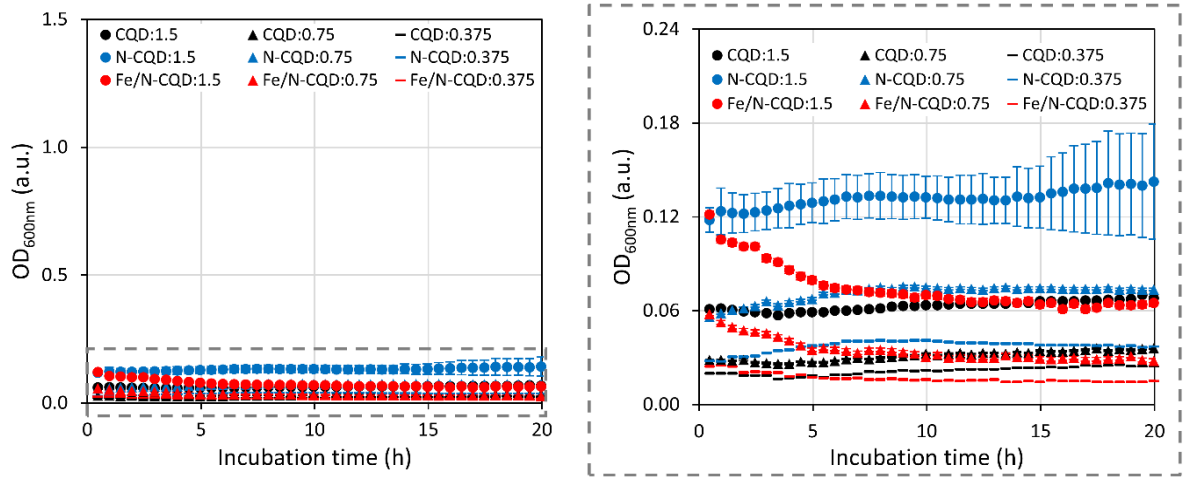

*Figure S2 Optical density measurements taken at 600 nm ($OD_{600nm}$) of the three greatest concentrations of CQDs without bacteria in MHB (37 °C, 20 h, continuous shaking). Data points are the average of a single experiment measured in triplicate (n = 3).*

# Final optical density values after growth ($OD_{final}$)

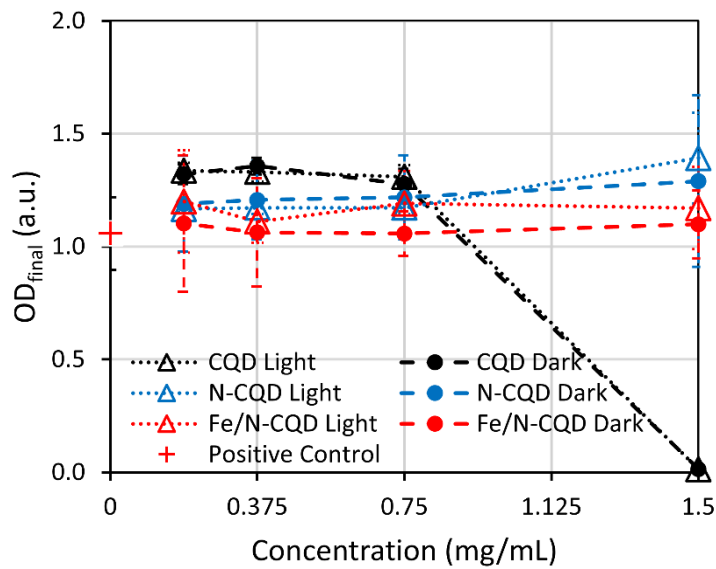

*Figure S3 Optical density measurements taken at the end of the bacteria growth experiment ($OD_{final}$). Data points are the mean and error bars indicate the standard deviation of the mean (n = 6).*

# Scanning Electron Microscopy

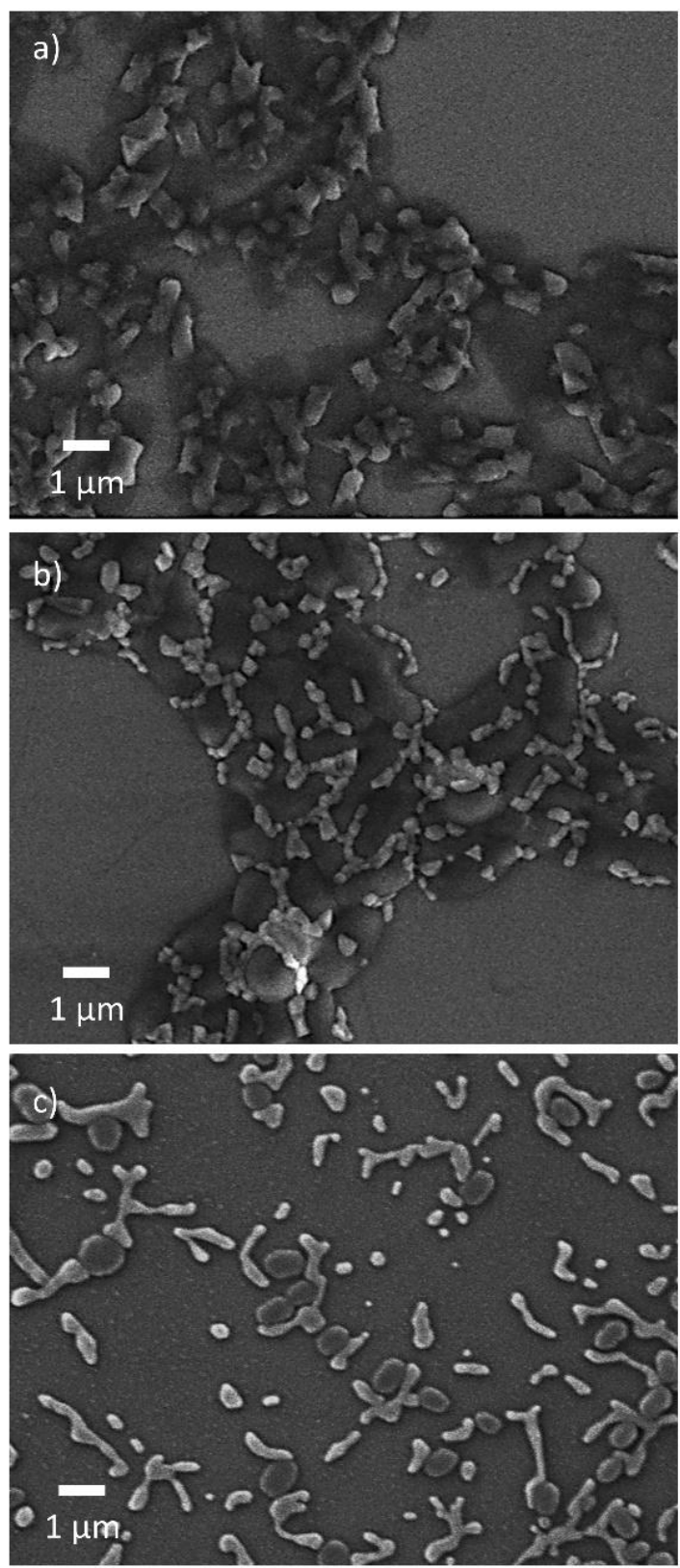

**Figure S4** *Wider field of view of the individual images used to create the composite image in main manuscript (**Figure 5**) showing a) CQDs, b) N-CQDs and c) Fe/N-CQDs.*